\documentclass[aps,amsmath,amssymb,superscriptaddress,prb,preprint]{revtex4-1}

\usepackage{graphicx}
\usepackage{bm}
\usepackage{color} 
\begin{document}

\title{Current-induced magnetization switching using electrically-insulating spin-torque generator}

\author{Hongyu An}
\affiliation{Department of Applied Physics and Physico-Informatics, Keio University, Yokohama 223-8522, Japan}

\author{Takeo Ohno}
\affiliation{WPI-Advanced Institute for Materials Research, Tohoku University, Sendai 980-8577, Japan}
\affiliation{PRESTO, Japan Science and Technology Agency, Kawaguchi, Saitama 332-0012, Japan}

\author{Yusuke Kanno} 
\affiliation{Department of Applied Physics and Physico-Informatics, Keio University, Yokohama 223-8522, Japan}

\author{Yuito Kageyama} 
\affiliation{Department of Applied Physics and Physico-Informatics, Keio University, Yokohama 223-8522, Japan}

\author{Yasuaki Monnai} 
\affiliation{Department of Applied Physics and Physico-Informatics, Keio University, Yokohama 223-8522, Japan}

\author{Hideyuki Maki} 
\affiliation{Department of Applied Physics and Physico-Informatics, Keio University, Yokohama 223-8522, Japan}
\affiliation{PRESTO, Japan Science and Technology Agency, Kawaguchi, Saitama 332-0012, Japan}

\author{Ji Shi} 
\affiliation{School of Materials and Chemical Technology, Tokyo Institute of Technology, Tokyo 152-8552, Japan}

\author{Kazuya Ando\footnote{Correspondence and requests for materials should be addressed to ando@appi.keio.ac.jp}}
\affiliation{Department of Applied Physics and Physico-Informatics, Keio University, Yokohama 223-8522, Japan}
\affiliation{PRESTO, Japan Science and Technology Agency, Kawaguchi, Saitama 332-0012, Japan}

\maketitle

\textbf{
Current-induced magnetization switching through spin-orbit torques (SOTs) is the fundamental building block of spin-orbitronics. The SOTs generally arise from the spin-orbit coupling of heavy metals. However, even in a heterostructure where a metallic magnet is sandwiched by two different insulators, a nonzero current-induced SOT is expected because of the broken inversion symmetry; an electrical insulator can be a spin-torque generator. Here, we demonstrate current-induced magnetization switching using an insulator. We show that oxygen incorporation into the most widely used spintronic material, Pt, turns the heavy metal into an electrically-insulating generator of the SOTs, enabling the electrical switching of perpendicular magnetization in a ferrimagnet sandwiched by electrically-insulating oxides. We further found that the SOTs generated from the Pt oxide can be controlled electrically through voltage-driven oxygen migration. These findings open a route towards energy-efficient, voltage-programmable spin-orbit devices based on solid-state switching of heavy metal oxidation.
}

An emerging direction in spintronics aims at discovering novel phenomena and functionalities originating from the spin-orbit coupling (SOC) in solid-state devices~\cite{RevModPhys.87.1213,spinHalldevices,manchon2015new,soumyanarayanan2016emergent}. Of particular recent scientific and technological interests are current-induced spin-orbit torques, which are responsible for the manipulation of magnetization in ultrathin ferromagnetic metals~\cite{gambardella2011current,miron2011perpendicular,miron2010current,liu2012spinScience,liu2012current,kim2013layer,yu2014switching,fan2014magnetization,wang2014spin,garello2013symmetry,fan2013observation,demasius2016enhanced}. The current-induced magnetization switching through the spin-orbit torques is generally observed in structures with broken inversion symmetry along the growth direction, such as an ultrathin ferromagnetic metal (FM) sandwiched between an oxide and a heavy metal (HM): oxide/FM/HM structures. In the oxide/FM/HM heterostructures, the spin-orbit torques have two components with different symmetries: damping-like and field-like spin-orbit torques~\cite{garello2013symmetry,kim2013layer}. Recent experimental and theoretical progress has revealed that both the bulk and interface SOC can result in the exertion of the damping-like and field-like spin-orbit torques~\cite{2012arXiv1204.4869M,haney2013current,pesin2012quantum}.

Since the spin-orbit torques due to the interface SOC originate from the broken symmetry of the heterostructure, it is natural to expect that a nonzero spin-orbit torque can be generated even in a heterostructure where a metallic magnet is sandwiched by two different insulators; an electrical insulator can be a spin-torque generator purely driven by the interface SOC. The electrically-insulating spin-torque generator will allow electrical switching of magnetization free from the energy dissipation of the charge current in the bulk of the spin-torque source, promising the development of energy-efficient spin-orbit devices. We also note that the underlying physics of the current-induced SOC effects is still in shape debate due to the fact that the spin-orbit torques originating from the bulk and interface SOC are difficult to disentangle in the metallic heterostructures. The electrically-insulating spin-torque generator provides a model system to study the spin-orbit effect purely arising from the interface SOC. However, despite of the recent vital progress in spin-orbitronics, the realization of magnetization manipulation using an insulating spin-torque generator remains elusive and challenging.

In this work, we demonstrate current-induced magnetization switching using an electrically-insulating spin-torque generator. We show that the widely used heavy metal, Pt, becomes an electrically-insulating generator of the spin-orbit torques after oxidation, enabling spin-torque magnetization switching in a heterostructure where a perpendicularly-magnetized ferrimagnetic metal is sandwiched by two insulating oxides: MgO and oxidized Pt. We found that even in the absence of any conducting HM, the oxygen-incorporated Pt, attached to a metallic magnet, generates a robust damping-like spin-orbit torque purely through interface spin-orbit scattering, which counters the conventional understandings of the spin-orbit torques, where the origin of the damping-like torque is primarily attributed to the bulk SOC. We further show that the relative strength of the damping-like and field-like torques changes systematically depending on the oxidation level of the Pt layer. This allows electrical tuning of the spin-orbit torques through voltage-driven O$^{2-}$ migration near the FM/oxygen-incorporated-Pt interface. These results provide crucial piece of information for revealing the underlying physics behind the current-induced spin-orbit torques as well as for the practical application of spin-orbit devices.

\bigskip\noindent
\textbf{Current-induced magnetization switching}

\begin{figure}[tb]
\includegraphics[scale=1]{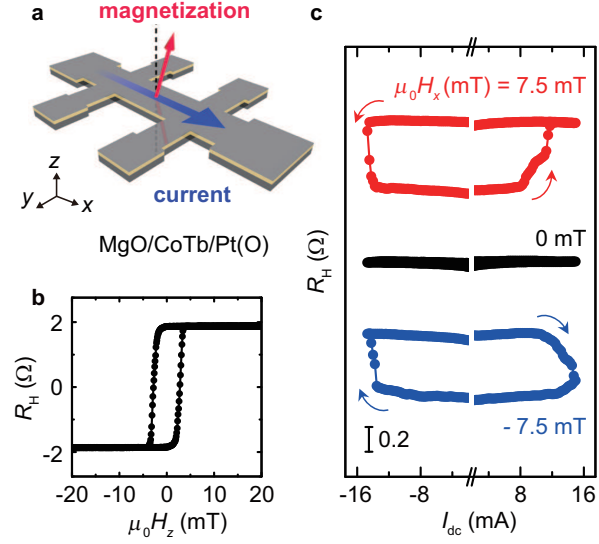}
\caption{
{\bfseries Current-induced magnetization switching.} \textbf{a}, Schematic of the MgO/CoTb/Pt(O) heterostructure used for the AHE measurements. The charge current $I_\text{dc}$ and external magnetic field were applied along the $x$ axis for the magnetization switching. \textbf{b}, The anomalous Hall resistance $R_\text{H}$ measured by varying the perpendicular magnetic field $\mu_0H_{z}$ for the MgO/CoTb/Pt(O) device. \textbf{c}, Current-induced magnetization switching curves for the MgO/CoTb/Pt(O) heterostructure, measured with different in-plane external magnetic field $\mu_0H_{x}$. }
\label{fig6} 
\end{figure}

The device for the current-induced magnetization switching using an electrically-insulating spin-torque generator is a MgO(1.4~nm)/CoTb(4.2~nm)/Pt(O)(8~nm)/substrate trilayer, capped with a Pt(1.7~nm) film, where the numbers in parentheses represent the thickness. Here, the atomic ratio of Tb in CoTb alloy was set as 0.25 and the thickness of the CoTb layer was set as 4.2~nm to obtain optimal perpendicular magnetic anisotropy on the Pt(O) layer~\cite{finley2016spin,han2017room}. The films were patterned into a 20 $\mu$m $\times$ 80 $\mu$m Hall bar shape (see Fig.~\ref{fig6}a). The oxygen-incorporated Pt, Pt(O), film was deposited on a thermally oxidized Si (SiO$_2$) substrate at room temperature by radio frequency (rf) magnetron sputtering (for details, see Methods and Supplementary Note~1). For the sputtering, argon and oxygen gases were introduced into the chamber and the amount of oxygen gas in the reactive mixture, $Q$, was set as 16\% for the current-induced magnetization switching experiment. A challenge for realizing the current-induced magnetization switching is to fabricate FM layer with perpendicular magnetic anisotropy on the Pt(O) film. We tried the commonly used Co and CoFeB as the FM layer. However, their magnetic anisotropy strongly depends on the interfacial conditions, which precludes to obtain perpendicular magnetic anisotropy on the Pt(O) film. Thus, we chose a transition metal-rare earth ferrimagnetic alloy CoTb with robust perpendicular magnetic anisotropy in the bulk as the FM layer for the current-induced magnetization switching~\cite{hansen1989magnetic,finley2016spin,han2017room}. Figure~\ref{fig6}b shows the anomalous Hall resistance (AHE) $R_\text{H}$ for the MgO/CoTb/Pt(O) film measured by sweeping an perpendicular magnetic field $\mu_0H_{z}$. The nearly square-shape magnetic hysteresis loop of $R_\text{H}$ demonstrates a good perpendicular magnetic anisotropy in the MgO/CoTb/Pt(O) film.

In the MgO/CoTb/Pt(O) trilayer, due to the broken inversion symmetry along the growth direction, a nonzero spin-orbit torque is expected when passing a charge current. Thus, we next measure $R_\text{H}$ as a function of an in-plane dc current $I_\text{dc}$. For the measurement, we applied an in-plane magnetic field $\mu_0H_{x}$ along the $x$ axis to break the rotational symmetry of the spin-orbit torque. As shown in Fig.~\ref{fig6}c, by applying a nonzero magnetic field of $\mu_0H_{x}$ = 7.5 mT, the current switches the magnetization of the CoTb layer between up and down directions. By reversing the direction of the magnetic field, the polarity of the magnetization switching is also reversed. When $\mu_0H_{x}=0$, the magnetization switching disappears. These features are consistent with the magnetization switching induced by the spin-orbit torque.

The above experimental result unambiguously demonstrates the current-induced magnetization switching in the heterostructure of ferrimagnetic CoTb sandwiched by two oxides: MgO and Pt(O). This result indicates that a nonzero spin-orbit torque is generated in the heterostructure, in spite of the fact that the applied charge current flows only in the ferrimagnetic CoTb layer; the current flow in the bulk of the Pt(O) layer can be neglected due to the much larger resistivity of Pt(O) (22917 $\mu\Omega$cm) than that of CoTb (74 $\mu\Omega$cm). In the MgO/CoTb/Pt(O) trilayer, since the ferrimagnetic layer is sandwiched between the electrically-insulating oxides, only the MgO/CoTb and CoTb/Pt(O) interfaces can be responsible for the spin-orbit-torque generation. Previous studies have shown that, by the Rashba effect, a MgO/FM interface primarily generates a field-like torque~\cite{akyol2015effect,ou2016origin}, which cannot be responsible for the current-induced switching in the heterostructure; in order to switch the magnetization, a sizable damping-like torque is required~\cite{liu2012spinScience}. We have confirmed the existence of a damping-like effective field in the heterostructure by measuring the second harmonic of the AHE resistance (see Supplementary Note~2). This result suggests that the CoTb/Pt(O) interface is responsible for the spin-torque generation. However, the origin responsible for this damping-like-torque generation is still unclear. Therefore, in the following, we focus on the spin-torque generation from the FM/Pt(O) interface.

\bigskip\noindent
\textbf{Generation of spin-orbit torques from oxygen incorporated Pt}

\begin{figure}[tb]
\includegraphics[scale=1]{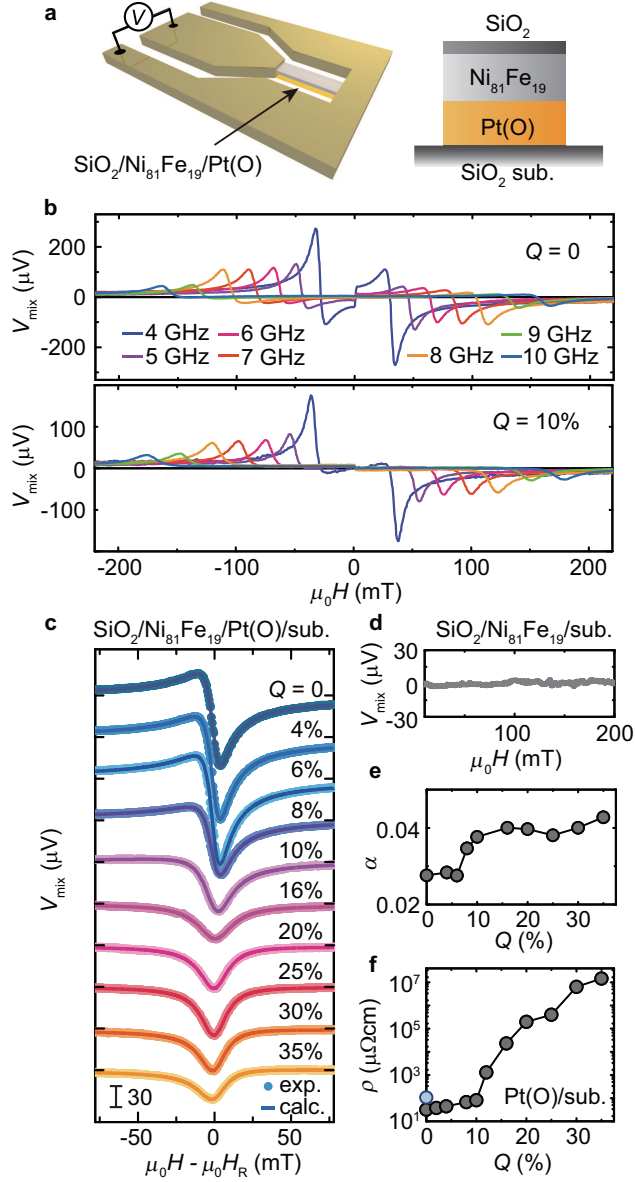}
\caption{
{\bfseries ST-FMR measurements.} \textbf{a}, Schematic of the SiO$_2$/Ni$_{81}$Fe$_{19}$/Pt(O) device for the ST-FMR measurements. \textbf{b}, ST-FMR spectra for the SiO$_2$/Ni$_{81}$Fe$_{19}$/Pt(O) devices by changing the rf current frequencies from 4 to 10 GHz, where $Q = 0$ and 10\%. \textbf{c}, ST-FMR spectra for the SiO$_2$/Ni$_{81}$Fe$_{19}$/Pt(O) devices at 7 GHz by changing $Q$ from 0 to 35\%. The solid circles are the experimental data and the solid curves are the fitting result using equation~(\ref{SandA}). The rf power of 24.7 dBm was applied for all the measurements. \textbf{d}, The ST-FMR spectrum for the SiO$_2$/Ni$_{81}$Fe$_{19}$ device at 7 GHz. \textbf{e}, $Q$ dependence of the magnetic damping constant $\alpha$, obtained from the rf frequency $f$ dependence of the ST-FMR spectral width $W$ using $W=(2\pi \alpha/\gamma)f +W_\text{ext} $, where $\gamma$ is the gyromagnetic ratio and $W_\text{ext}$ is the extrinsic contribution to the spectral width. \textbf{f}, $Q$ dependence of the electrical resistivity $\rho$ of Pt(O) films. The blue solid circle shows the electrical resistivity of a Ni$_{81}$Fe$_{19}$ film.} 
\label{fig1} 
\end{figure}

In order to systematically study the generation of the spin-orbit torques from oxygen-incorporated  Pt, we measure the spin-torque ferromagnetic resonance (ST-FMR) for Ni$_{81}$Fe$_{19}$/Pt(O) bilayers with $Q$ in a range of 0 to 35\%. As shown in Fig.~\ref{fig1}a, after the deposition of a Pt(O) film, a 6-nm-thick Ni$_{81}$Fe$_{19}$ layer was deposited on the Pt(O) layer and a 4-nm-thick SiO$_2$ capping layer was used to protect the Ni$_{81}$Fe$_{19}$ layer from oxidation. The SiO$_2$/Ni$_{81}$Fe$_{19}$/Pt(O) films were patterned into rectangular strips with 10 $\mu$m width and 150 $\mu$m length using the photolithography and lift-off techniques. For the ST-FMR measurement, an rf charge current was applied along the longitudinal direction and an in-plane external magnetic field $\bf H$ was applied with an angle of 45$^\circ$ from the longitudinal direction of the device. In the device, the rf charge current generates the spin-orbit torques as well as an Oersted field, driving magnetization precession in the Ni$_{81}$Fe$_{19}$ layer. The magnetization precession results in an oscillation of the resistance due to the anisotropic magnetoresistance in the Ni$_{81}$Fe$_{19}$ layer, which can be measured through the mixing dc voltage $V_\text{mix}$ by using a bias tee~\cite{liu2011spin}. The mixing voltage $V_\text{mix}$ can be expressed as~\cite{liu2011spin, zhang2015role}
\begin{equation}
V_\text{mix}=S\frac{W^2}{(\mu_0H-\mu_0H_\text{FMR})^2+W^2}+A\frac{W(\mu_0H-\mu_0H_\text{FMR})}{(\mu_0H-\mu_0H_\text{FMR})^2+W^2},\label{SandA}
\end{equation}
where $S$, $A$, $W$ and $\mu_0H_\text{FMR}$ are the magnitude of the symmetric component, the magnitude of the antisymmetric component, the spectral width, and the FMR field, respectively. Here, the symmetric component is proportional to the damping-like effective field $H_\text{DL}$ and the antisymmetric component is due to the sum of the Oersted field $H_\text{Oe}$ and the field-like effective field $H_\text{FL}$~\cite{PhysRevB.92.064426}.

Figure~\ref{fig1}b shows the ST-FMR spectra $V_\text{mix}$ for the SiO$_2/$Ni$_{81}$Fe$_{19}$/Pt(O) devices with $Q=0$ and 10\% measured at a frequency range of 4-10 GHz. As can be seen, by reversing the external magnetic field $\bf H$ direction, the sign of $V_\text{mix}$ also changes correspondingly, as expected for the voltage generation induced by the ST-FMR. In order to systematically investigate the influence of the oxidation level in the Pt(O) films on the spin-torque generation, we have measured the ST-FMR for the SiO$_2$/Ni$_{81}$Fe$_{19}$/Pt(O) films with different $Q$ from 0 to 35\% as shown in Fig~\ref{fig1}c. Notably, the spectral shape of $V_\text{mix}$ changes drastically by increasing the amount of the oxygen gas flow $Q$; the ratio between symmetric and antisymmetric components, $S/A$, increases with increasing $Q$. This result shows that the spin-torque generation efficiency in the SiO$_2$/Ni$_{81}$Fe$_{19}$/Pt(O) device is strongly affected by the oxidation level of the Pt(O) layer. We have confirmed that the voltage signal disappears in a SiO$_2$/Ni$_{81}$Fe$_{19}$ film fabricated on a SiO$_2$ substrate as shown in Fig.~\ref{fig1}d. The absence of the voltage signal in the SiO$_2$/Ni$_{81}$Fe$_{19}$ film supports that the change in the spin-orbit torques generated from the Pt(O) layer is responsible for the change in the ST-FMR spectral shape in the SiO$_2$/Ni$_{81}$Fe$_{19}$/Pt(O) films.

The observed change in the ST-FMR spectral shape is associated with the interface SOC in the SiO$_2$/Ni$_{81}$Fe$_{19}$/Pt(O) films. In Fig.~\ref{fig1}e, we show the magnetic damping constant $\alpha$ determined from frequency dependence of the ST-FMR spectral width for the SiO$_2$/Ni$_{81}$Fe$_{19}$/Pt(O) film with various $Q$. The magnetic damping $\alpha$ is clearly enhanced by increasing the oxidation level of the Pt(O) layer, especially for $Q>10$\%. In the SiO$_2$/Ni$_{81}$Fe$_{19}$/Pt(O) film, the magnetic damping is dominated by the dissipation of the angular momentum induced by the spin pumping; the spin pumping emits a spin current from the Ni$_{81}$Fe$_{19}$ layer, and the absorption of the spin current outside the Ni$_{81}$Fe$_{19}$ layer deprives the magnetization of the angular momentum, giving rise to the additional magnetic damping~\cite{Tserkovnyak1,Mizukami} . However, the magnetic damping enhanced by the oxygen incorporation in the SiO$_2$/Ni$_{81}$Fe$_{19}$/Pt(O) film cannot be attributed to the spin absorption in the interior of the Pt(O) layer. When $Q>10$\%, the electrical resistivity of the Pt(O) layer increases drastically with $Q$ (see Fig.~\ref{fig1}f), and thus the spin pumping into the bulk of the Pt(O) layer is strongly suppressed. With the spin absorption in the bulk of the Pt(O) layer ruled out as a mechanism behind the enhancement of the magnetic damping, the only possible mechanism that agrees with the experimental observation is the spin absorption at the Ni$_{81}$Fe$_{19}$/Pt(O) interface. With strong interface SOC, the Ni$_{81}$Fe$_{19}$/Pt(O) interface can be an efficient spin absorber because of the fast spin relaxation due to the spin-momentum coupling of the Rashba state and/or efficient spin-flip scattering due to the SOC at the interface. Thus, the enhancement of the magnetic damping, as well as the change in the ST-FMR spectral shape, suggests that the strength of the SOC at the Ni$_{81}$Fe$_{19}$/Pt(O) interface is enhanced by increasing the oxidation level of the Pt(O) layer.

\begin{figure}[tb]
\includegraphics[scale=1]{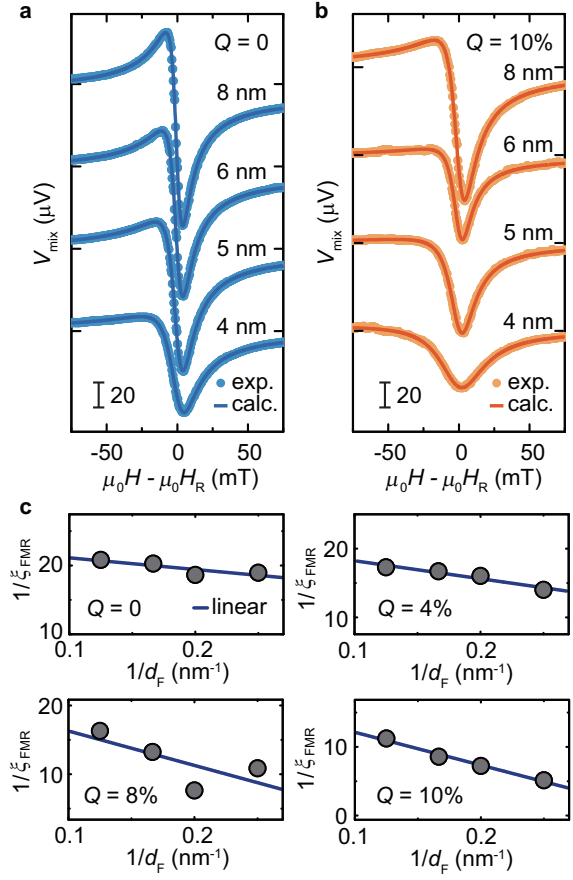}
\caption{
{\bfseries Thickness dependence of ST-FMR.} ST-FMR spectra for the SiO$_2$/Ni$_{81}$Fe$_{19}$/Pt(O) devices at 7 GHz when $Q$ is (\textbf{a}) 0 and (\textbf{b}) 10\%. The Ni$_{81}$Fe$_{19}$-layer thickness $d_\text{F}$ was changed from 4 to 8 nm.  \textbf{c}, Inverse of the FMR spin-torque generation efficiency $1/\xi_\text{FMR}$ as a function of 1/$d_\text{F}$ for $Q=0$, 4\%, 8\%, and 10\%. The solid circles are the experimental data and the solid lines are the linear fit to the data. }
\label{fig2} 
\end{figure}

\begin{figure}[tb]
\includegraphics[scale=1]{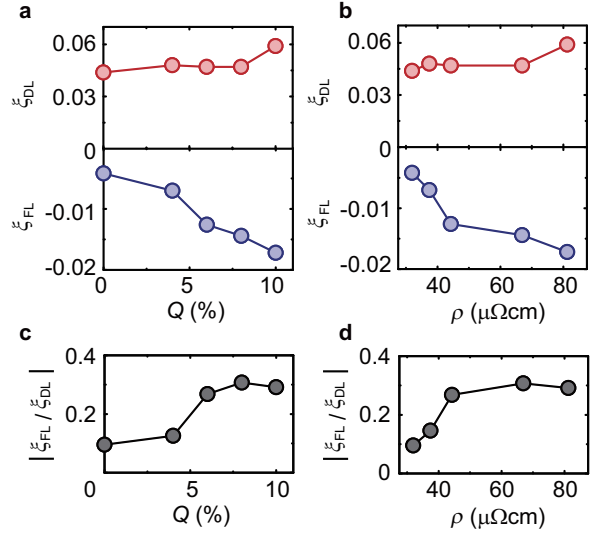}
\caption{
{\bfseries Spin-torque generation efficiencies when $Q$ below 10\%.} \textbf{a}, $Q$ dependence of the damping-like $\xi_\text{DL}$ and field-like $\xi_\text{FL}$ spin-torque generation efficiencies. \textbf{b}, Pt(O)-layer-resistivity $\rho$ dependence of $\xi_\text{DL}$ and $\xi_\text{FL}$. \textbf{c}, $Q$ dependence of the ratio between $\xi_\text{FL}$ and $\xi_\text{DL}$.  \textbf{d}, Pt(O)-layer-resistivity $\rho$ dependence of the ratio between $\xi_\text{FL}$ and $\xi_\text{DL}$.}
\label{fig3} 
\end{figure}

In order to quantitatively investigate the influence of the oxidation level of the Pt(O) layer on the generation efficiency of the spin-orbit torques, we first determine the damping-like $\xi_\text{DL}$ and field-like $\xi_\text{FL}$ spin-torque efficiencies for the moderately oxidized Pt(O) films ($Q \leq 10\%$) by measuring Ni$_{81}$Fe$_{19}$ layer thickness $d_\text{F}$ dependence of the ST-FMR spectra. By increasing $Q$ from 0 to 10\% the electrical resistivity of the Pt(O) films only increases from 32 $\mu\Omega$cm to 81 $\mu\Omega$cm as shown in Fig.~\ref{fig1}f, which is still smaller than that of the Ni$_{81}$Fe$_{19}$ film (106 $\mu\Omega$cm, see the blue solid circle in Fig.~\ref{fig1}f). Therefore, the Oersted field created by the charge current flowing in the Pt(O) layer cannot be neglected in the SiO$_2$/Ni$_{81}$Fe$_{19}$/Pt(O) film. In the presence of the Oersted field, the FMR spin-torque generation efficiency obtained from the resonance lineshape,
\begin{equation}
\xi_\text{FMR}=\frac{S}{A}\frac{e\mu_0M_\text{s} d_\text{F}d_\text{N}}{\hbar}\sqrt{1+\frac{\mu_0M_\text{s}}{\mu_0H_\text{FMR}}},\label{xiFMR}
\end{equation}
is related to the damping-like $\xi_\text{DL}$ and field-like $\xi_\text{FL}$ spin-torque efficiencies, under the assumption that $\xi_\text{DL(FL)}$ does not have a strong dependence on $d_\text{F}$ in the range examined, as~\cite{PhysRevB.92.064426}
\begin{equation}
\frac{1}{\xi_\text{FMR}}=\frac{1}{\xi_\text{DL}} \left(1+\frac{\hbar}{e}\frac{\xi_\text{FL}}{\mu_0M_\text{s}d_\text{F}d_\text{N}} \right), \label{thicknesseqa}
\end{equation}
where ${\xi _{{\rm{DL(FL)}}}} =  {\mu_0{M_{\rm{s}}}} {d_{{\rm{F}}}}{H_{{\rm{DL(FL)}}}}\left( {2e/\hbar } \right)/{j_{\rm{c}}}^\text{Pt(O)}$. 
Here, $d_\text{F}$ and $d_\text{N}$ are the thicknesses of the Ni$_{81}$Fe$_{19}$ layer and Pt(O) layer, respectively. ${j_{\rm{c}}}^\text{Pt(O)}$ is the charge current density in the Pt(O) layer and $\mu_0 M_\text{s}$ is the saturation magnetization. Figures~\ref{fig2}a and \ref{fig2}b show the ST-FMR spectra for the SiO$_2$/Ni$_{81}$Fe$_{19}$/Pt(O) films with different $d_\text{F}$ in a range of 4-8 nm for $Q = 0$ and $Q = 10\%$. The ratio between symmetric and antisymmetric components $S/A$ in both spectra increases by decreasing $d_\text{F}$. From the 1/$d_\text{F}$ dependence of 1/ $\xi_\text{FMR}$ shown in Fig.~\ref{fig2}c with equation~(\ref{thicknesseqa}), $\xi_\text{DL}$ and $\xi_\text{FL}$ are  determined as shown in Fig.~\ref{fig3}a. As can be seen, both $\xi_\text{DL}$ and $\xi_\text{FL}$ increase with $Q$; by increasing $Q$ from 0 to 10\%,  $\xi_\text{DL}$ increases from 0.044 to 0.059 and  $\xi_\text{FL}$ increases from $-0.0042$ to $-0.017$. The minus sign of  $\xi_\text{FL}$ indicates that the field-like effective field is opposite to the Oersted field. In Fig.~\ref{fig3}b, $\xi_\text{DL}$ and $\xi_\text{FL}$ are replotted as a function of the electrical resistivity $\rho$ of the Pt(O) layer. Here, in a system where the damping-like torque is purely generated by the bulk spin Hall effect, the damping-like spin-torque efficiency $\xi_\text{DL}$ is equivalent to the spin Hall angle $\theta_\text{SHE}$. Since $\theta_\text{SHE}$ is known to scale with the electrical resistivity~\cite{0034-4885-78-12-124501} $\rho$, the nonmonotonic change of $\xi_\text{DL}$ with $\rho$ shown in Fig.~\ref{fig3}b suggests that the bulk spin Hall effect is not the only source of the observed spin-orbit torques in the SiO$_2$/Ni$_{81}$Fe$_{19}$/Pt(O) films with $Q \leq 10\%$. We also found that the relative magnitude of the field-like torque to the damping-like torque, $\xi_\text{FL}/\xi_\text{DL}=H_\text{FL}/H_\text{DL}$, increases with $Q$ or $\rho$ as shown in Figs.~\ref{fig3}c and \ref{fig3}d.

\begin{figure}[bt]
\includegraphics[scale=1]{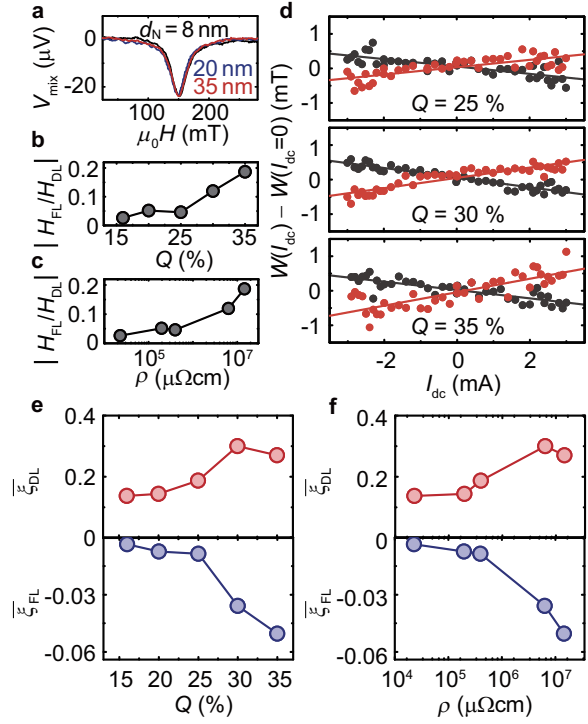}
\caption{
{\bfseries Spin-torque generation efficiencies when $Q$ above 16\%.} \textbf{a}, Pt(O)-layer-thickness $d_\text{N}$ dependence of the ST-FMR spectra for the SiO$_2$/Ni$_{81}$Fe$_{19}$/Pt(O) devices at 7 GHz with $Q= 20$\%. \textbf{b}, $Q$ dependence of the ratio between the damping-like $H_\text{DL}$ and field-like $H_\text{FL}$ effective fields. \textbf{c}, Pt(O)-layer-resistivity $\rho$ dependence of the ratio between $H_\text{DL}$ and $H_\text{FL}$.  \textbf{d}, The change of linewidth $W$ of the ST-FMR spectrum as a function of the applied dc current $I_\text{dc}$ for different $Q$. \textbf{e}, $Q$ dependence of {$\bar{\xi}_\text{DL}$ and $\bar{\xi}_\text{FL}$}. \textbf{f}, $\rho$ dependence of $\bar{\xi}_\text{DL}$ and $\bar{\xi}_\text{FL}$. }
\label{fig4} 
\end{figure}

By further increasing the oxidation level of the Pt(O) layer, the charge and spin transport in the SiO$_2$/Ni$_{81}$Fe$_{19}$/Pt(O) film change drastically. For $Q \geq 16\%$, the electrical resistivity of the Pt(O) films is larger than $2.3 \times10^4$ $\mu\Omega$cm, which is more than two-orders of magnitude larger than the resistivity of the Ni$_{81}$Fe$_{19}$ layer (see also Fig.~\ref{fig1}f). This indicates that the flow of the applied rf charge current in the Pt(O) layer can be neglected for $Q \geq 16\%$. Consequently, the Oersted field arising from the charge current in the Pt(O) layer is negligible in the SiO$_2/$Ni$_{81}$Fe$_{19}$/Pt(O) films. This is evidenced in the ST-FMR spectra measured for different Pt(O) layer thicknesses shown in Fig.~\ref{fig4}a. As can be seen, the ST-FMR spectral shape barely changes by varying the Pt(O) layer thickness from 8 to 35 nm for $Q=20$\%. Since the antisymmetric component of $V_\text{mix}$ due to the Oersted field $H_\text{Oe}$ = $d_\text{N}{j_{\rm{c}}}^\text{Pt(O)}/2$ is proportional to the thickness of the Pt(O) layer $d_\text{N}$, the spectral shape is independent of $d_\text{N}$ only when the charge current density ${j_{\rm{c}}}^\text{Pt(O)}$ in the Pt(O) layer is negligible. The negligible ${j_{\rm{c}}}^\text{Pt(O)}$ in the Pt(O) layer for $Q\geq16$\% is also confirmed by a numerical calculation using CST microwave studio.

In the absence of the charge current in the interior of the Pt(O) layer, the spin-orbit torques due to the bulk spin Hall effect, as well as the torque due to the Oersted field, is negligible. Nevertheless, we observed the clear ST-FMR signals for the SiO$_2$/Ni$_{81}$Fe$_{19}$/Pt(O) devices with $Q \geq 16\%$ as shown in Fig.~\ref{fig1}c. This result demonstrates that the current-induced FMR is driven by the spin-orbit toques purely generated at the Ni$_{81}$Fe$_{19}$/Pt(O) interface. Here, the absence of the Oersted field in the device precludes the determination of the spin-torque efficiencies, $\xi_\text{DL}$ and $\xi_\text{FL}$, with the self-calibrated equations~(\ref{xiFMR}) and (\ref{thicknesseqa}), used for $Q\leq10$\%, where the strength of the spin-orbit torques from the charge current is measured relative to the torque from the Oersted field. This is because $\xi_\text{DL}$ and $\xi_\text{FL}$ are defined as the generation efficiency of the spin-orbit torques from the charge current density $j_\text{c}^\text{Pt(O)}$ flowing in the Pt(O) layer; $\xi_\text{DL}$ and $\xi_\text{FL}$ are undefined in the absence of ${j_{\rm{c}}^\text{Pt(O)}}$. Under the negligible Oersted field, instead of the spin-torque efficiencies $\xi_\text{DL}$ and $\xi_\text{FL}$, the relative magnitude of the field-like effective field $H_\text{FL}$ to the damping-like effective field $H_\text{DL}$ can be obtained directly from the ST-FMR spectral shape as 
\begin{equation}
\frac{{{H_{{\rm{FL}}}}}}{{{H_{{\rm{DL}}}}}} = \frac{A}{S}{\left( {1 + \frac{{{\mu _0}{M_{\rm{s}}}}}{{{\mu _0}{H_{{\rm{FMR}}}}}}} \right)^{ - 1/2}}. \label{HDLHFLa}
\end{equation}
In Fig.~\ref{fig4}b, we show $Q$ dependence of $H_\text{FL}$/$H_\text{DL}$ obtained using equation~(\ref{HDLHFLa}). $H_\text{FL}$/$H_\text{DL}$ is also plotted as a function of the Pt(O)-layer resistivity $\rho$ in Fig.~\ref{fig4}c. Although these results show significant increase of $H_\text{FL}$/$H_\text{DL}$ with the oxidation of the Pt(O) layer, the damping-like torque is about an order of magnitude larger than the field-like torque in the trilayer, where the Ni$_{81}$Fe$_{19}$ layer is sandwiched by the two insulating oxides. This observation is well consistent with the result in our current-induced magnetization switching experiment that the damping-like torque is responsible for the switching, but contrary to the prediction of the spin-orbit torques generated by the interface Rashba effect; previous experimental and theoretical results suggest that the Rashba effect primarily generates the field-like torque~\cite{2012arXiv1204.4869M,haney2013current,pesin2012quantum}. Our results therefore require a new source of the spin-orbit torques, other than the bulk spin Hall effect and interface Rashba effect, in the highly-oxidized SiO$_2$/Ni$_{81}$Fe$_{19}$/Pt(O) films.

A possible origin of the spin-orbit torques in the highly-oxidized SiO$_2$/Ni$_{81}$Fe$_{19}$/Pt(O) films is interfacial spin-dependent scattering~\cite{wang2016giant,PhysRevB.94.104420}. With strong interface SOC, the interfacial spin-orbit scattering creates spin currents that can flow away from the interface, which enter the ferromagnetic layer. This mechanism can exert torques on the magnetization through the spin-transfer mechanism, allowing the spin-orbit torques to have a strong damping-like component~\cite{PhysRevB.94.104420}, which is consistent with our observation. Another possible mechanism is the intrinsic spin-orbit torque~\cite{kurebayashi2014antidamping}. In a two-dimensional Rashba system, a spin-orbit effective field induces spin rotation, which results in non-equilibrium out-of-plane spin polarization of carriers that are exchange coupled to the in-plane magnetization. The out-of-plane spin density induces the damping-like torque on the magnetization~\cite{kurebayashi2014antidamping}. Thus, with the assumption that the Rashba SOC is present at the Ni$_{81}$Fe$_{19}$/Pt(O) interface, the observed damping-like torque can be explained by the intrinsic mechanism in the highly-oxidized SiO$_2$/Ni$_{81}$Fe$_{19}$/Pt(O) films.

To characterize the generation efficiency of the damping-like and field-like torques for the SiO$_2$/Ni$_{81}$Fe$_{19}$/Pt(O) device with $Q \geq 16\%$, we measured the ST-FMR with applying a dc charge current $I_\text{dc}$. In the SiO$_2$/Ni$_{81}$Fe$_{19}$/Pt(O) device, the damping-like torque generated by the dc charge current $I_\text{dc}$ effectively changes the magnetic damping $\alpha$ of the Ni$_{81}$Fe$_{19}$ layer, or the FMR spectral width $W$, as shown in Fig.~\ref{fig4}d. The damping modulation enables to determine the conversion efficiency from the applied charge current $I_\text{dc}$ into the damping-like effective field $H_\text{DL}$. Here, note that for $Q \geq 16\%$, the applied charge current flows only in the Ni$_{81}$Fe$_{19}$ layer, and thus the damping modulation allows to determine the generation efficiency of the damping-like torque from the charge current density $j_\text{c}^\text{Py}$ flowing in the Ni$_{81}$Fe$_{19}$ layer: ${\bar{\xi} _{{\rm{DL}}}} =  {\mu_0{M_{\rm{s}}}} {d_{{\rm{F}}}}{H_{{\rm{DL}}}}\left( {2e/\hbar } \right)/{j_{\rm{c}}^\text{Py}}$, which is different from ${\xi _{{\rm{DL}}}} =  {\mu_0{M_{\rm{s}}}} {d_{{\rm{F}}}}{H_{{\rm{DL}}}}\left( {2e/\hbar } \right)/{j_{\rm{c}}^\text{Pt(O)}}$ obtained for $Q \leq 10\%$. Thus, although ${\bar{\xi} _{{\rm{DL}}}}$ characterizes the variation of the generation efficiency of the damping-like torque with the oxidation level of the electrically-insulating Pt(O) layer, ${\bar{\xi} _{{\rm{DL}}}}$ cannot be compared directly with ${\xi _{{\rm{DL}}}}$ obtained for $Q \leq 10\%$. We also note that the damping-like effective field $H_\text{DL}$ for $Q \geq 16\%$ is generated by the charge current flowing at the Ni$_{81}$Fe$_{19}$/Pt(O) interface. This indicates that ${\bar{\xi} _{{\rm{DL}}}}$ is not the actual spin-torque efficiency due to the interfacial spin-dependent scattering. Nevertheless, as shown in Fig.~\ref{fig4}e, we found large ${\bar{\xi} _{{\rm{DL}}}}$ for $Q \geq 16\%$. This result demonstrates that the spin-orbit scattering at the Ni$_{81}$Fe$_{19}$/Pt(O) interface provides an efficient way for generating the spin-orbit torques, which are large enough to manipulate the magnetization of the ferromagnetic layer. The ${\bar{\xi} _{{\rm{DL}}}}$ result also allows to determine the generation efficiency, ${\bar{\xi} _{{\rm{FL}}}}$, of the field-like torque from $j_\text{c}^\text{Py}$ using ${\bar{\xi} _{{\rm{FL}}}}/{\bar{\xi} _{{\rm{DL}}}}=H_\text{FL}$/$H_\text{DL}$ with the measured values of $H_\text{FL}$/$H_\text{DL}$ shown in Figs.~\ref{fig4}b and \ref{fig4}c. As shown in Figs.~\ref{fig4}e and \ref{fig4}f, the field-like torque efficiency ${\bar{\xi} _{{\rm{FL}}}}$ increases drastically with the oxidation level of the Pt(O) layer, illustrating a crucial role of the oxidation of the Pt layer for the field-like torque generation.

\bigskip\noindent
\textbf{Voltage control of spin-orbit torques}

\begin{figure*}[bt]
\includegraphics[scale=1]{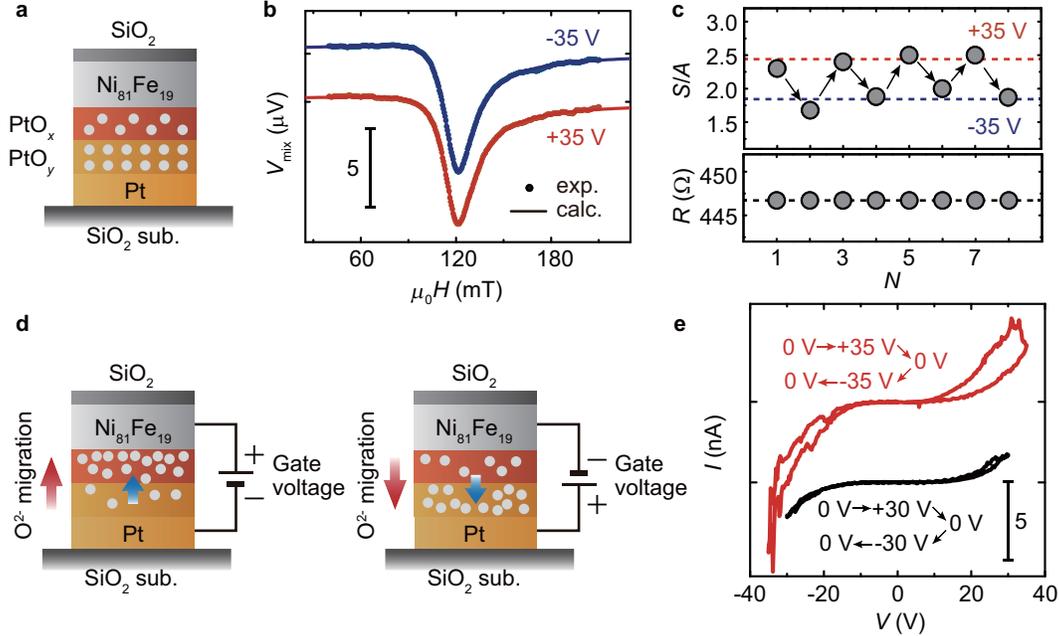}
\caption{
{\bfseries Spin-torque generation controlled by O$^{2-}$ migration.} \textbf{a}, Schematic of the heterostructure used for the O$^{2-}$ migration and ST-FMR measurement. The gray solid circles represent oxygen ions. \textbf{b}, Typical ST-FMR spectra measured after removing the applied voltages of $\pm 35$ V. The solid circles are the experimental data and the solid curves are the fitting result using equation~(\ref{SandA}). The offset of the curves in vertical direction was shifted for comparison. \textbf{c}, The magnitude of the $S/A$ ratio obtained by fitting the corresponding ST-FMR spectra, where $N$ represents the cycle index. The ST-FMR were measured for the Ni$_{81}$Fe$_{19}$/Pt(O)/Pt device after the application of the gate voltage of $+35$ V ($N=1,3,5,7$) or $-35$ V ($N=2,4,6,8$). The in-plane electrical resistance $R$ of the Ni$_{81}$Fe$_{19}$ layer in the Ni$_{81}$Fe$_{19}$/PtO$_x$/PtO$_y$/Pt device measured after removing the applied voltages of $\pm 35$ V is plotted correspondingly. \textbf{d}, Schematic of the experimental setup for the application of the gate voltages used to drive the O$^{2-}$ migration. The O$^{2-}$ migrates towards the Ni$_{81}$Fe$_{19}$/Pt(O) interface for the application of the positive gate voltage (left), whereas the negative gate voltage drives O$^{2-}$ away from the Ni$_{81}$Fe$_{19}$/Pt(O) interface (right). \textbf{e}, Typical current-voltage ($I$-$V$) curves measured across the Ni$_{81}$Fe$_{19}$/PtO$_x$/PtO$_y$/Pt junction. The offset of the curves in vertical direction was shifted for comparison.}
\label{fig5} 
\end{figure*}

The above study unambiguously reveals the significant influence of the oxidation level of the Ni$_{81}$Fe$_{19}$/Pt(O) bilayer on the spin-orbit torque generation and provides a probable approach to tune the spin-orbit torques through voltage driven O$^{2-}$ migration near the Ni$_{81}$Fe$_{19}$/Pt(O) interface. The voltage driven O$^{2-}$ migration has been well studied in a wide range of oxides so far~\cite{waser2009redox,yang2013memristive,bauer2015magneto,yang2008memristive,yang2009family,lee2011fast,yang2012demand,doi:10.1063/1.4929512}. However, a crucial problem is that the oxygen concentration in the magnetron-sputtered Pt(O) films is too low to be utilized for the O$^{2-}$ migration even when we fabricated the Pt(O) film by using the oxygen flow $Q$ of 100$\%$ during the deposition. To solve this problem, we fabricated a highly resistive PtO$_x$/PtO$_y$ bilayer using the reactive sputtering and the oxygen plasma treatment as shown in Fig.~\ref{fig5}a. The 7-nm-thick PtO$_x$ layer was deposited by using the oxygen flow $Q$ of 100$\%$ on the PtO$_y$ layer, where the PtO$_y$ layer was a magnetron-sputtered Pt layer oxidized by oxygen particle irradiation at the beginning~\cite{panda2001anisotropic,economou2008fast,ohno2015resistive,ohno2015and,cheng2016preparation}; firstly, a 7.5-nm-thick Pt layer was deposited on a SiO$_2$ substrate, and then accelerated oxygen particles, extracted from the conventional oxygen plasma, passing through a biased electrode, were used to oxidize the Pt film. This oxygen plasma treatment leads to a formation of PtO$_y$ with a depth of around 3.5 nm from the surface. The remaining non-oxidized Pt was utilized as the bottom electrode for the application of a gate voltage.  The oxidized Pt layer created by the oxygen particle irradiation guarantees the sufficiently high oxygen concentration and the PtO$_x$/PtO$_y$ bilayer system provides an oxygen concentration gradient which favors the O$^{2-}$ migration~\cite{yang2008memristive,yang2009family,lee2011fast}. On the PtO$_x$/PtO$_y$ bilayer, a 6-nm-thick Ni$_{81}$Fe$_{19}$ layer with a 4-nm-thick SiO$_2$ capping layer was deposited, which serves as both the FM layer for the ST-FMR measurement and the top electrode for the application of the gate voltage. For the Ni$_{81}$Fe$_{19}$/PtO$_x$/PtO$_y$/Pt (Ni$_{81}$Fe$_{19}$/Pt(O)/Pt) device, we first applied the gate voltage between the top Ni$_{81}$Fe$_{19}$ and bottom Pt layers and then measured the ST-FMR after removing the gate voltage.

Figure~\ref{fig5}b shows typical ST-FMR spectra measured after removing the gate voltage of $\pm 35$ V from the Ni$_{81}$Fe$_{19}$/Pt(O)/Pt device. The spectral shape of the ST-FMR changes by reversing the polarity of the gate voltage. Figure~\ref{fig5}c summarizes the ratio between symmetric and antisymmetric component $S/A$ obtained by fitting the ST-FMR spectra, which demonstrates reversible switching of the $S/A$ ratio induced by the voltage application.

This reversible manipulation is best interpreted in terms of the switching of the spin-orbit torque generation due to internal O$^{2-}$ or vacancies migration through nanoionic transport as illustrated in Fig.~\ref{fig5}d. When the positive gate voltage (0 V$\rightarrow$ $+35$ V$\rightarrow$ 0 V) was applied, the O$^{2-}$ migrates towards the Ni$_{81}$Fe$_{19}$/Pt(O) interface, leading to an increase of the $S/A$ ratio due to the high oxygen incorporation in the PtO$_x$ layer. In contrast, the application of the negative gate voltage (0 V$\rightarrow$ $-35$ V$\rightarrow$ 0 V), which drives O$^{2-}$ away from the Ni$_{81}$Fe$_{19}$/Pt(O) interface, results in a decrease of the $S/A$ ratio. This is consistent with the oxidation level dependence of the ST-FMR spectral shape shown in Fig.~\ref{fig1}c, where the $S/A$ ratio increases by increasing the oxidation level in the Ni$_{81}$Fe$_{19}$/Pt(O) bilayer.

The reversible switching of the ST-FMR spectral shape was observed only when the large voltages ($\pm 35$ V) were applied. This result indicates a nonvolatile characteristic when the gate large voltages ($\pm 35$ V) is applied, whereas relative low gate voltages result in volatile characteristic. These differences appear in the current-voltage ($I$-$V$) curves shown in Fig.~\ref{fig5}e. When the gate voltages up to 30 V is applied, the curve follows an almost same history, which corresponds to a volatile O$^{2-}$ migration. On the other hand, by applying the voltages up to 35 V, a resistive switching characteristic is observed in the $I$-$V$ curve, indicating a nonvolatile O$^{2-}$ migration.

During the reversible switching of the spin-orbit torque generation, although a possible O$^{2-}$ migration into the Ni$_{81}$Fe$_{19}$ layer can also change the $S/A$ ratio, the oxidation of the Ni$_{81}$Fe$_{19}$ layer cannot be responsible for the change of the ST-FMR spectra. Since the gate voltage is applied along the perpendicular direction of the device (see Fig.~\ref{fig5}d), if the O$^{2-}$ migrates into the Ni$_{81}$Fe$_{19}$ layer, the electrical resistance along the in-plane direction of the Ni$_{81}$Fe$_{19}$ layer should change. As shown in Fig.~\ref{fig5}c, however, we observed no change in the in-plane electrical resistance of the Ni$_{81}$Fe$_{19}$ layer in the Ni$_{81}$Fe$_{19}$/Pt(O)/Pt device after the application of $\pm 35$ V. This result indicates that the oxidation of the Ni$_{81}$Fe$_{19}$ layer is negligible. A possible O$^{2-}$ migration into the bottom non-oxidized Pt layer also plays a minor role in the voltage-induced change of the ST-FMR spectra. Although the spin-orbit torques generated by the spin Hall effect in the bottom Pt layer can be neglected due to the presence of the insulating Pt(O) layer in the Ni$_{81}$Fe$_{19}$/Pt(O)/Pt device, an Oersted field created by the rf charge current flowing in the bottom Pt layer may contribute to the ST-FMR signal. However, by applying a negative voltage, if the O$^{2-}$ migrates into the bottom Pt layer, the effective thickness of the non-oxidized Pt layer should be decreased and thus the Oersted field should also be decreased. This is contrary to the fact that the $S/A$ switching shown in Fig.~\ref{fig5}c can be reproduced only when the negative(positive) voltage increases(decreases) the Oersted field. Therefore, the voltage-induced change of the ST-FMR spectral shape can be attributed to the O$^{2-}$ migration within the highly-resistive PtO$_x$/PtO$_y$ bilayer; the O$^{2-}$ migration in the PtO$_x$/PtO$_y$ bilayer changes the oxidation level at the Ni$_{81}$Fe$_{19}$/PtO$_x$ interface, which enables to tune the spin-orbit torque generation.

\bigskip\noindent
\textbf{Summary}

In summary, we have demonstrated current-induced magnetization switching using an electrically-insulating spin-torque generator. By turning Pt into an electrical insulator through oxygen incorporation, we show that even without the bulk spin Hall effect, still a robust damping-like torque can be generated through interface spin-orbit scattering. The spin-torque generation at FM/Pt(O) interfaces allows to switch the magnetization in a perpendicularly magnetized ferrimagnet sandwiched between insulating oxides. We note that in the MgO/CoTb/Pt(O) device used in the present study, a large current is necessary to switch the perpendicular magnetization. This is because the most of the applied charge current flows in the bulk of the thick CoTb layer in spite of the fact that the spin-orbit torque responsible for the magnetization switching is purely generated by the interface current. Thus, the switching current can be decreased significantly by fabricating a thinner perpendicularly-magnetized film; the insulating spin-torque generator promises a way to minimize the switching energy of spin-based magnetic memories because of the absence of the energy dissipation due to the current flow in the bulk of the spin-torque generator. Furthermore, we show that the spin-orbit effect can be tuned electrically by changing the oxidation level near the FM/Pt(O) interface through the voltage-driven O$^{2-}$ migration in the oxygen-incorporated Pt layer. This finding is different from the previous report that the voltage-driven O$^{2-}$ migration was used to oxidize a FM layer to tune the magnetic properties~\cite{bauer2015magneto}, since we use the voltage-driven O$^{2-}$ migration to directly tune the spin torque generator instead of the FM layer. As the energy dissipation in the bulk can be avoided, our findings pave a way towards the energy-efficient, voltage-programmable spintronic applications.

\clearpage

\clearpage

\bigskip\noindent
\textbf{Methods}\\
\textbf{Sample preparation.} The films were deposited on thermally oxidized Si substrates (SiO$_2$) by rf magnetron sputtering at room temperature. Before the deposition, the base pressure in the chamber  was better than 1$\times$10$^{-6}$ Pa and the deposition pressure was 0.2 Pa. The CoTb, Ni$_{81}$Fe$_{19}$ and SiO$_2$ films were deposited by applying argon gas with flow of 10 sccm. For the Pt(O) deposition, argon and oxygen gases were introduced into the chamber and the amount of oxygen gas in the reactive mixture, $Q$, was changed in order to change the oxygen concentration in the Pt(O) films. The film thicknesses were controlled by the deposition time with a pre-calibrated deposition rate. For the fabrication of the devices used in magnetization switching experiment, the substrates were patterned into a 20 $\mu$m$\times80$ $\mu$m Hall bar shape by standard photolithography before the deposition, and lift-off technique was used to take off the rest part of the films after the deposition. For the fabrication of the devices used in the ST-FMR experiment, the substrates were patterned into 10 $\mu$m$\times150$ $\mu$m rectangular shape. For the fabrication of the devices used in the experiment of the voltage-tuning of the oxidation level at the Ni$_{81}$Fe$_{19}$/Pt(O) interface, a Pt film was oxidized by irradiation with accelerated oxygen particles passing through a 30 W-biased electrode with cylindrical holes. The source power for an inductively coupled oxygen plasma generation was rf 500 W. To investigate the microstructure of Pt(O), X-ray diffraction (XRD) and X-ray reflectivity (XRR) measurements were conducted with a Bruker D8 Discover diffractometer by applying Cu K$_\alpha$ radiation. Single Pt(O) blanket films were fabricated by changing oxygen flow for the measurements. The electrical resistivity of Pt(O) and Ni$_{81}$Fe$_{19}$ films was measured using the conventional four-probe method. All measurements were conducted at room temperature.

\clearpage

\bigskip\noindent
Correspondence and requests for materials should be addressed to K.A. (ando@appi.keio.ac.jp)\\

\bigskip\noindent
\textbf{Acknowledgements}\\
This work was supported by JSPS KAKENHI Grant Numbers 26220604, 26103004, PRESTO-JST ``Innovative nano-electronics through interdisciplinary collaboration among material, device and system layers" Grant Number 13415036, JPMJPR1323 and JPMJPR1325, the Mizuho Foundation for the Promotion of Sciences, JSPS Core-to-Core Program, and Spintronics Research Network of Japan (Spin-RNJ). H.A. acknowledges the support from the JSPS Fellowship (No.~P17066). T.O. thanks Prof. S. Samukawa for fruitful discussion on the plasma technique.

\bigskip\noindent
\textbf{Additional information}\\
The authors declare no competing financial interests.

\bigskip\noindent
\textbf{Author contributions}\\
H.A. and T.O. fabricated devices and collected the data. H.A., T.O., Y. Kanno, Y. Kageyama and Y.M. analyzed the data. K.A., T.O., H.M. and J.S. designed the experiments and developed the explanation. K.A., T.O. and H.A. wrote the manuscript. All authors discussed results and reviewed the manuscript.

\clearpage

\end{document}